\documentclass{svproc}
\usepackage{url}

\usepackage{graphicx}

\usepackage{microtype}
\usepackage{subfigure}
\usepackage{float}
\usepackage{booktabs} 

\usepackage{amssymb}
\usepackage{hyperref}
\PassOptionsToPackage{hyphens}{url}
\usepackage{lipsum}
\usepackage{physics}
\usepackage{tikz}
\usetikzlibrary{automata,positioning}

\newcommand{\be}{\begin{eqnarray}}
\newcommand{\ee}{\end{eqnarray}}

\begin{document}
\mainmatter             
\title{Quantum Circuits for Quantum Convolutions:\\A Quantum Convolutional Autoencoder}
\titlerunning{A Quantum Convolutional Autoencoder}

\author{Javier~Orduz\inst{1}
\and
Pablo~Rivas\inst{2}
\and
Erich~Baker\inst{3}
}
\authorrunning{J. Orduz et. al.} 
\tocauthor{
J. Orduz, 
P. Rivas, 
E. Baker
}
\institute{
School of Engineering and Computer Science \\
Department of Computer Science \\
Baylor University, Texas, USA\\
$^1$\email{Javier\_OrduzDucuara@baylor.edu}
$^2$\email{Pablo\_Rivas@baylor.edu}
$^3$\email{Erich\_Baker@baylor.edu}
}
\maketitle
\begin{abstract}
Quantum machine learning deals with leveraging quantum theory 
with classic machine learning algorithms. Current research efforts 
study the advantages of using quantum mechanics or quantum 
information theory to accelerate learning time or convergence. 
Other efforts study data transformations in the quantum information 
space to evaluate robustness and performance boosts. This paper 
focuses on processing input data using randomized quantum circuits 
that act as quantum convolutions producing new representations that 
can be used in a convolutional network. Experimental results suggest 
that the performance is comparable to classic convolutional neural 
networks, and in some instances, using quantum convolutions can 
accelerate convergence.
\keywords{Quantum Computing, Convolutional Autoencoder, Quantum Machine Learning}
\end{abstract}

\section{Introduction \label{sec:int}}
Neural network architectures based on convolutional operations are a trendy machine learning tool \cite{voulodimos2018deep,aloysius2017review,henning2014image}. These kinds of models offer certain versatility when the learned filters are transferred into or out of the network \cite{rai2020review}. Beyond the apparent filtering capabilities, convolutional neural networks (CNNs) have other interesting properties and architectures that have produced many exciting applications, and variants \cite{zhang2019graph,zhiqiang2017review,aloysius2017review}. This research is based on a classic autoencoder CNN architecture with convolutional and pooling layers. 

An autoencoder (AE) is considered an unsupervised learning model that reconstructs the input signal using a neural network \cite{dong2018review}. AEs are notably known for some of their successful versions, including the Variational Autoencoder (VAE) \cite{wetzel2017unsupervised}, and the denoising AE \cite{sagha2017stacked,rivas2019deep}. Convolutional AEs, in particular, have been proven to be very robust in learning representations of the data, usually compressed, while at the same time retaining much of the information \cite{guo2017deep}. In this research, we focus on an AE for image-related tasks, learning representations leveraging \textbf{quantum computing}, and replacing the first layer of the model. 

The field of quantum computing has grown recently, providing new means to calculate, represent, and read information \cite{biamonte2017quantum}. A recent study shows quantum machine learning can be considered a kernel transformation in classic machine learning \cite{schuld2021quantum}. Therefore, here we further explore the idea of using a quantum circuit to process images and then feed them to a convolutional autoencoder, which we call a \emph{quanvolutional autoencoder} \cite{rivas2021quanvolutional}. This work is also motivated by others who explored a quanvolutional network for classification purposes \cite{henderson2019quanvolutional}. 

The rest of the paper is organized as follows. Section 2 describes state of the art in quantum machine learning. Section 3 describes the proposed quanvolutional autoencoder architecture. Section 4 describes the results, and a discussion is presented in Section 5, along with our conclusions.

\section{Background \label{sec:bac}}

Quantum computation (QC) has permeated different computation areas; in particular, machine learning is a tangible subject to explore with quantum algorithms. Scientific community 
explores different tools and potential applications 
tackled with quantum algorithms and concepts related to Quantum Mechanics \cite{griffiths2018introduction}. 
On the other hand, Machine Learning (ML) techniques and 
algorithms, such as Supervised and Unsupervised, 
have been implemented in some python platforms. 
This intersection of ML and QC is called Quantum Machine Learning \cite{biamonte2017quantum,Schuld_2014,wittek2014quantum}.

One of the most interesting applications in machine learning models, in particular, autoencoders
\cite{Goodfellow-et-al-2016}.
This is a kind of artificial neural network belonging to the 
unsupervised learning, and it is useful to reduce the 
dimensions. The purpose of this document is to show some 
quantum advantages, several applications, in the 
Quanvolutional Neural Networks, which is model explored by 
\cite{henderson2019quanvolutional}. 

\begin{figure}[b!]\centering
    \begin{tikzpicture}[shorten >=1pt,node distance=2.5cm,on grid,auto] 
       \node[state] (i)   {$i$}; 
       \node[state] (c) [right=of i] {$c$}; 
       \node[state] (r) [right=of c] {$r$}; 
        \path[->] 
        (i) edge  node {encoder} (c)
        (c) edge  node  {decoder} (r);
    \end{tikzpicture}
    \caption{Basic autoencoder, where $i$ refers to the input, $c$ is the code, and  $r$ is the output.}
    \label{fig:auto}
\end{figure}
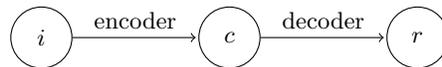

Autoencoders are generally used in tasks associated with information compression, and summarization \cite{theis2017lossy,yousefi2017text}, and generally speaking in dimensionality reduction \cite{wang2016auto,rivas2019deep}. 
Fig. \ref{fig:auto} depicts the basic representation of an autoencoder as a function with parameters that need to be learned efficiently. 
AEs are unsupervised models that aim to take an input, find intermediate linear and non-linear transformations, and reproducing or reconstructing the input at the output layer. Autoencoders usually minimize some reconstruction loss function such as the mean squared error or categorical cross-entropy reconstruction losses \cite{aloysius2017review}. The model presented in this work is a convolutional autoencoder whose architecture we discuss next.

\section{Autoencoder Architecture \label{sec:aut}}
\begin{figure}[ht!]
    \centering
    \includegraphics[width=\textwidth]{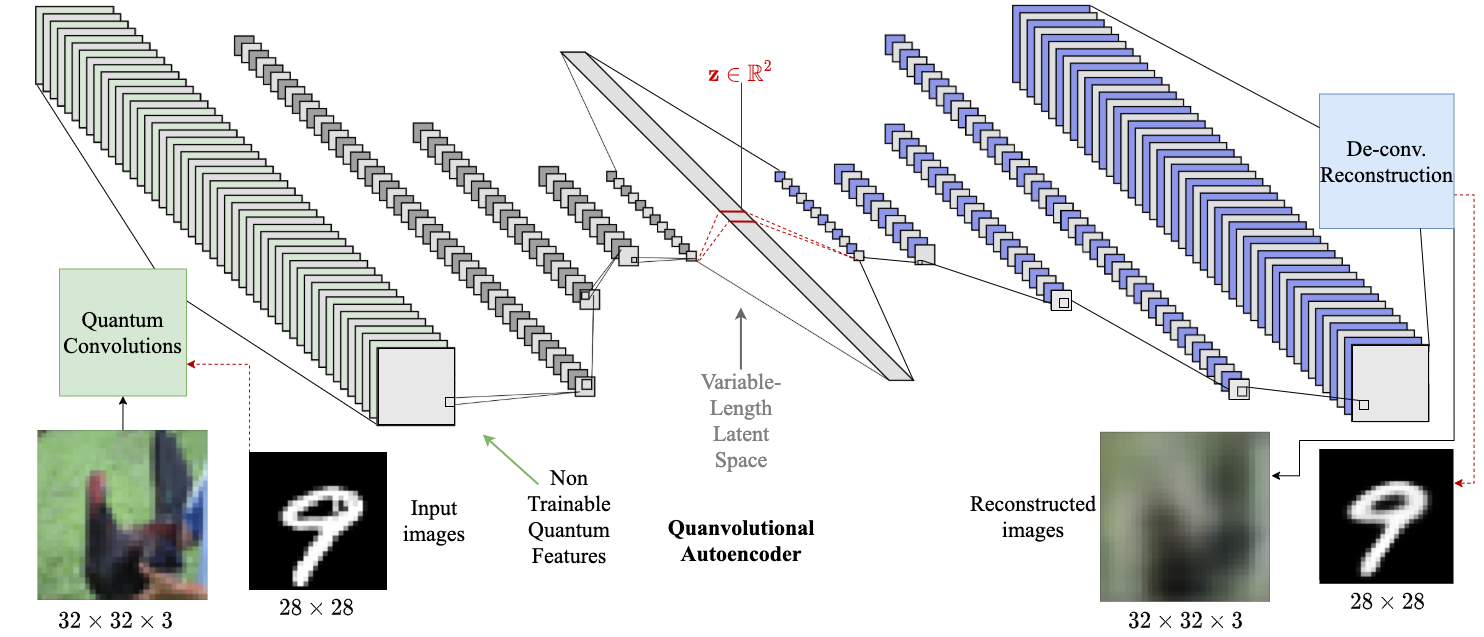}
    \caption{{\em Top:} Quanvolutional autoencoder architecture, where the quantum filtering (quantum convolutions) are described as quantum circuits, as shown in Fig. \ref{fig:15Q}. The quantum layer is non-trainable by the network. \emph{Bottom:} Classic convolutional autoencoder with all its layers trainable. Note that the dense layer in the center of the AEs can also vary, producing a two-dimensional latent space or a larger space.}
    \label{fig:architecture}
\end{figure}

We proposed to use the autoencoder architecture shown in Fig. \ref{fig:architecture}. 
\begin{figure}[ht!]\centering
    \includegraphics[width=0.7\textwidth]{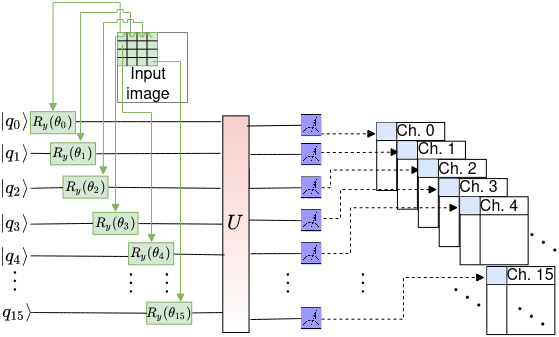}
    \caption{The quantum circuit for the autoencoder considering 16 qubits for a $4 \times 4$ quantum convolutional filter.}
    \label{fig:15Q}
\end{figure}
At the top and bottom of the figure, we have two similar configurations which differ in the first convolutional layer. The traditional autoencoder at the bottom of Fig. \ref{fig:architecture} shows an autoencoder that has the traditional convolutional, pooling, de-convolutional, and upsampling layers. However, the top of the figure shows an alternative configuration that uses quantum-based filters. These filters are implemented as quantum circuits that can produce an interesting and alternative way of representing images in lower-dimensional filters. Note, however, that in this particular configuration, we are using a non-trainable set of quantum-based filters. To this kind of architecture, we refer to as \textbf{Quanvolutional Autoencoder}. The name is inspired by the work of Henderson et al. \cite{henderson2019quanvolutional} on quanvolutional neural networks. Although a quantum version of the autoencoder has already been proposed by Romero et al. in 2017 \cite{romero2017quantum}, such model and the one presented here are substantially different and come from different motivations; our approach, in contrast, uses randomized quantum circuits as image convolutions.

Fig. \ref{fig:15Q} shows $|q_i^{}\rangle$ as the qubits on a $4\times 4$ region, each element of this region is associated to the same numbers of qubits initialized in ground state after we apply rotations, which are parametrized to angle $\theta$ and scaled by $\pi$, to embed the information in the qubits. This rotation is given by,
\be\label{eq:rotation}
R_{y}^{} (\theta) = e^{-i\theta \frac{\sigma_y^{}}{2}}_{}=
\begin{pmatrix}
\cos{\frac{\theta}{2}}&-\sin{\frac{\theta}{2}}\\
\sin{\frac{\theta}{2}}&\cos{\frac{\theta}{2}}\\
\end{pmatrix},
\ee
where eq.~\eqref{eq:rotation} represents the single qubit $Y$ rotation, $\sigma_y$ is the Pauli matrix, and $U$ is a random circuit, this part was implemented using PennyLane \cite{bergholm2020pennylane}. 

The quantum filtering layer is not available for training; this layer can be pre-computed prior to the training of the network to reduce the computation time. Otherwise, the quantum convolutional filtering process can be executed during mini-batch training on demand. The experimental design of this architecture is tested and compared to a classic approach where the entire network is trainable, as we discuss next.

\section{Experiments and Results \label{sec:exp}}

The experiments were carried out using a standard GPU-based system running Python 3 with PennyLane and TensorFlow libraries. We divided the experiment into two different ones with different datasets, MNIST and CIFAR-10. The objective of each experiment is to test the autoencoder's ability on a standard simple image dataset (MNIST) and a more complex, color-based one (CIFAR-10). We want to study the reconstruction ability of each model, inspect its latent space and its rate of convergence for both a quantum or a classic convolutional layer.

\subsection{MNIST}

\begin{figure}[ht!]\centering
    \begin{tabular}{c}
        \includegraphics[width=0.45\textwidth]{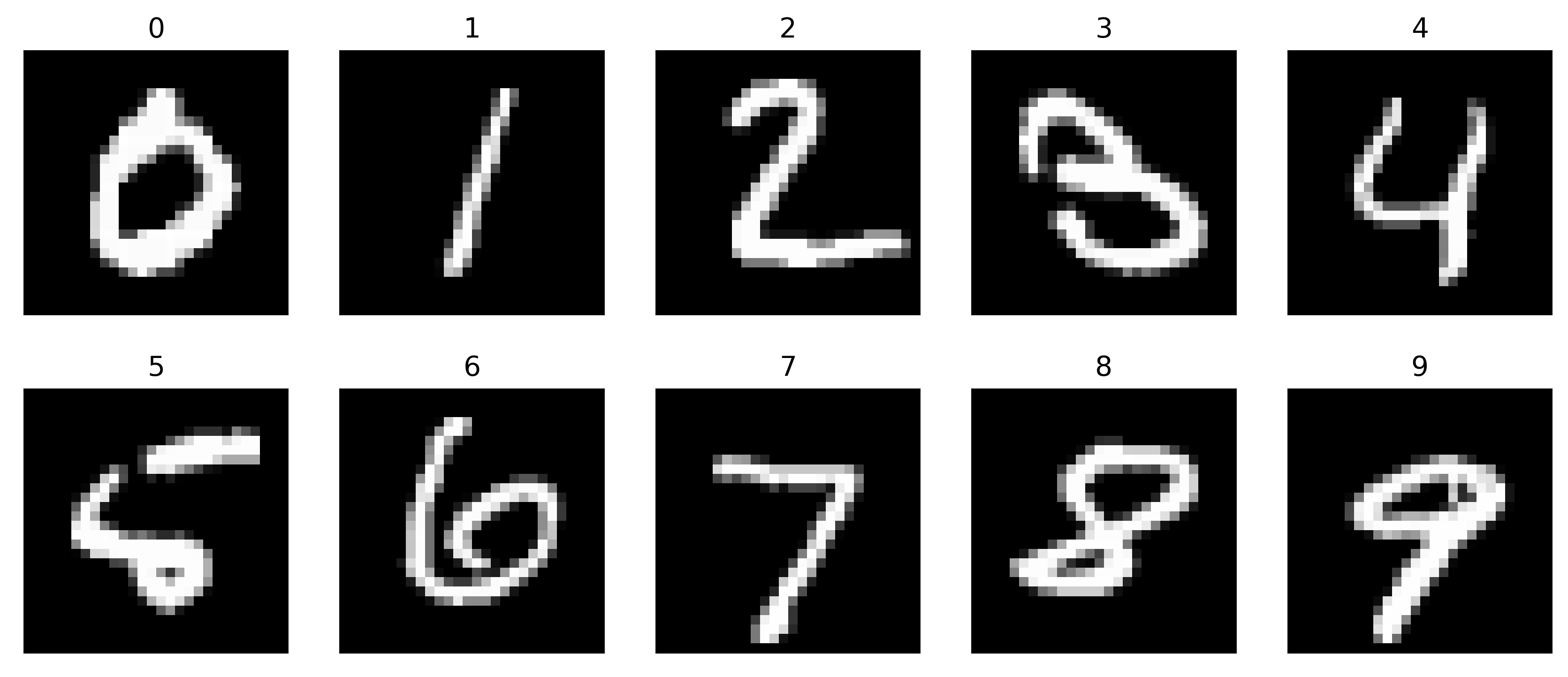} \\ (a) 
    \end{tabular}
    \begin{tabular}{cc}
        \includegraphics[width=0.45\textwidth]{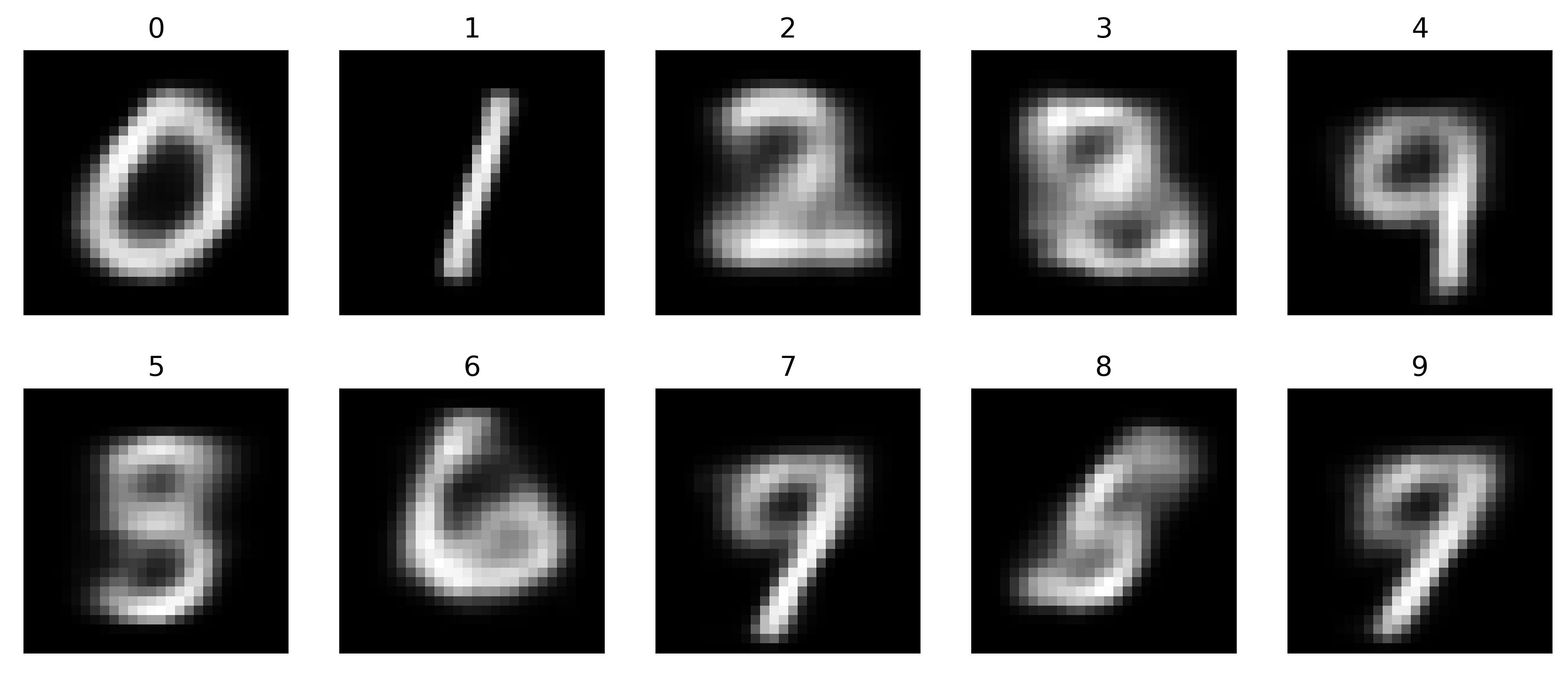} &
        \includegraphics[width=0.45\textwidth]{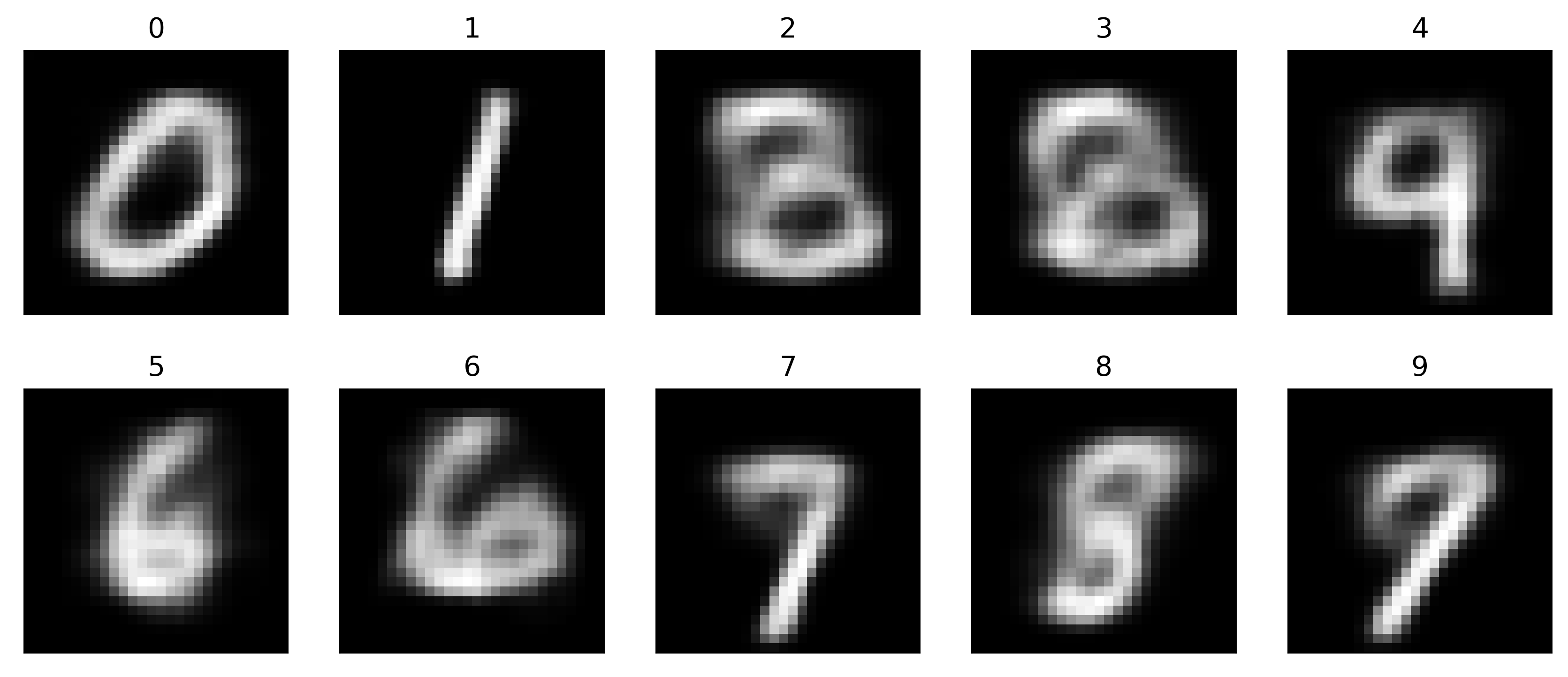} \\ (b) & (c) \\
        \includegraphics[width=0.45\textwidth]{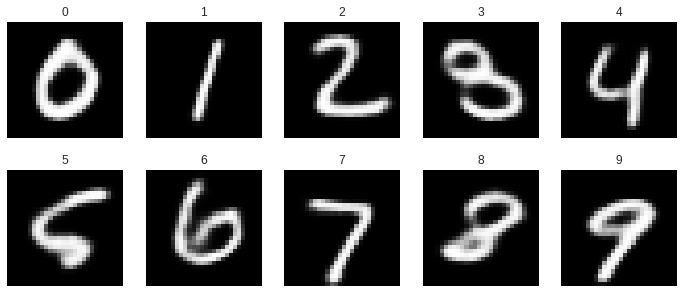} &
        \includegraphics[width=0.45\textwidth]{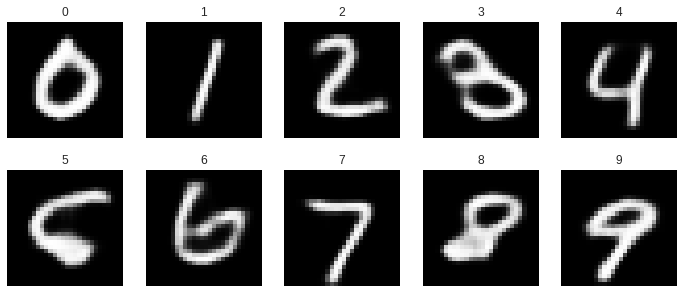} \\ (d) & (e) 
    \end{tabular}
    \caption{MNIST data and reconstruction results from different latent spaces. (a) Original. (b) Classic Convolutional AE, $\mathbf{z}\in \mathbb{R}^2$. (c) Quanvolutional AE, $\mathbf{z}\in \mathbb{R}^2$. (d) Classic Convolutional AE, $\mathbf{z}\in \mathbb{R}^{64}$. (e) Quanvolutional AE, $\mathbf{z}\in \mathbb{R}^{64}$.}
    \label{fig:mnist}
\end{figure}

The MNIST dataset \cite{lecun2010mnist} consists of grayscale images of size $28 \times 28$. The number of images used for training the AEs is $60,000$ and $10,000$ for testing. For this problem, we used the architecture described in \cite{rivas2021quanvolutional}. The results of the reconstruction using both the classic and quantum-based methodologies are shown in Fig. \ref{fig:mnist}. 
In (a) we have a sample of the original digits in the test set, then in (b) we have the classic convolutional AE approach in reconstruction while the quantum-based one is shown in (c). Clearly, the reconstruction abilities of the network are comparable. This initial test goes down to a latent space in $\mathbf{z} \in \mathbb{R}^2$; therefore, the reconstruction results appear to have considerable room for improvement. However, a second experiment was conducted where the autoencoders were set to find a latent space in $\mathbf{z} \in \mathbb{R}^{64}$, which yields the reconstruction results in Fig. \ref{fig:mnist} (d) and (e) for the classic and quantum approach, respectively. Evidently, these results in this larger latent space are much better in terms of reconstruction and denoising abilities. A visual comparison between both methods suggests that they have comparable reconstruction performance. 

\begin{figure}[ht!]
    \centering
    \begin{tabular}{cc}
        \includegraphics[width=0.5\textwidth]{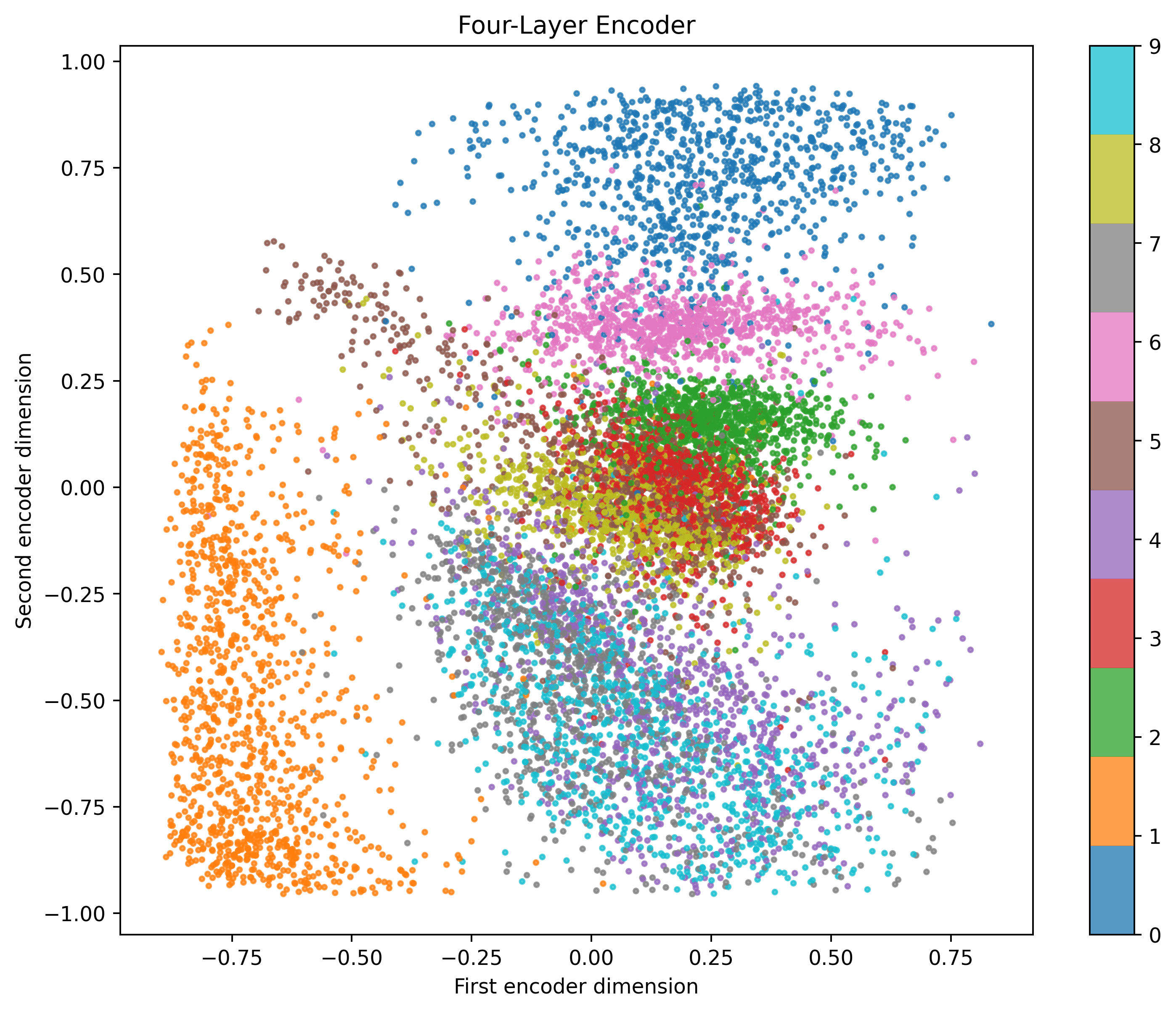} &
        \includegraphics[width=0.5\textwidth]{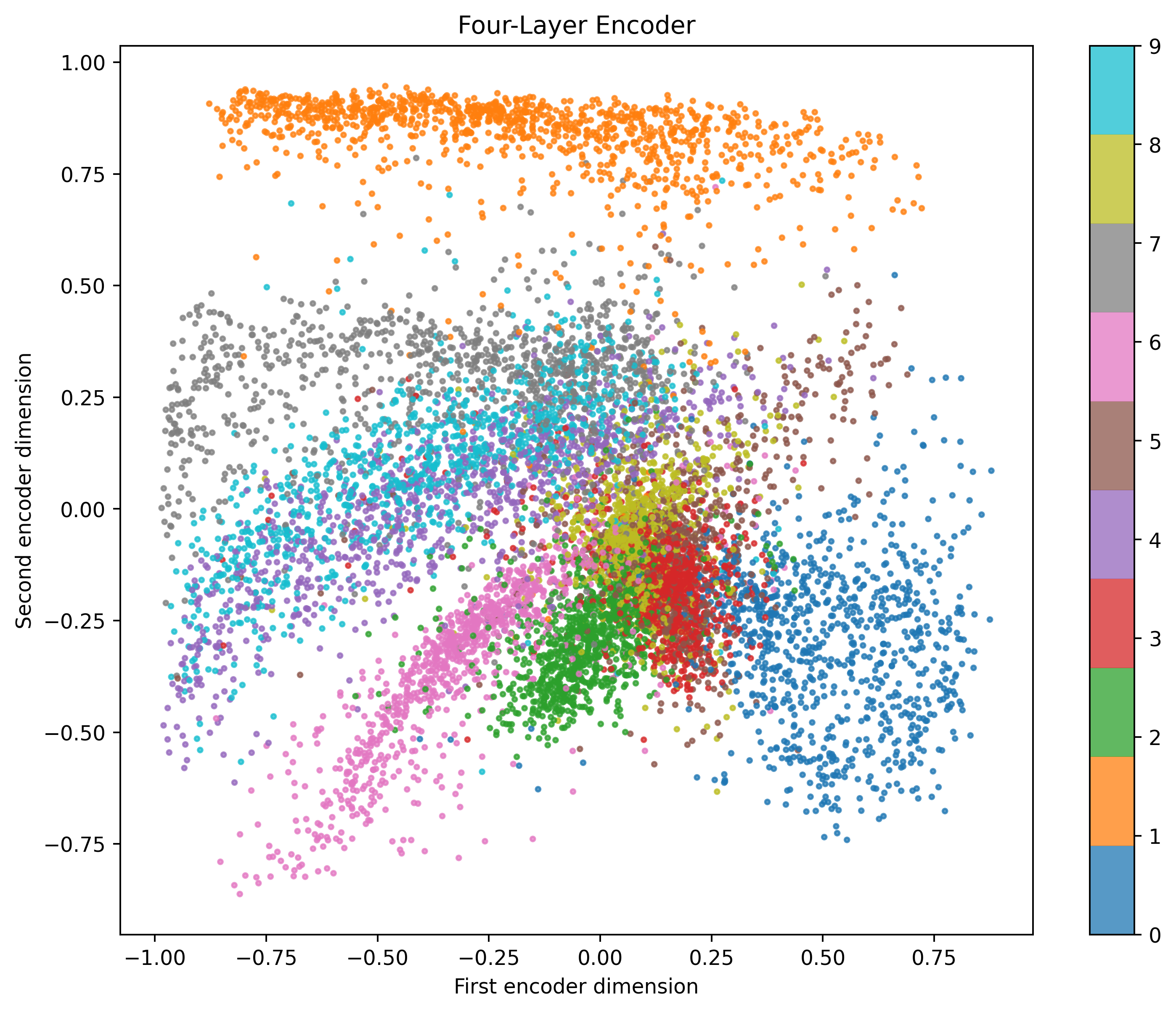} \\ (a) & (b) \\
        \includegraphics[width=0.5\textwidth]{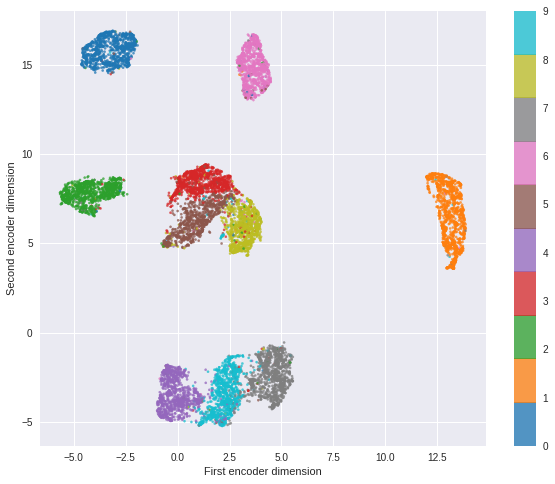} &
        \includegraphics[width=0.5\textwidth]{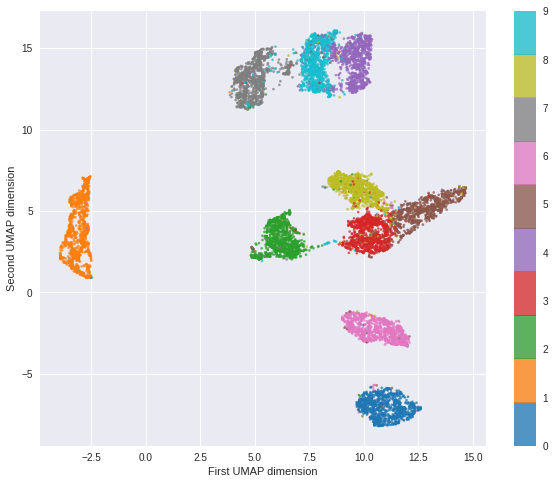} \\ (c) & (d) 
    \end{tabular}
    \caption{Two-dimensional representation of the latent space. (a) and (c) display classic AE while (b) and (d) depict the quantum-based.}
    \label{fig:mnistlatent}
\end{figure}

Projecting the test set on the learned latent space is one interesting way of examining what representations are being learned. To this end, we present Fig. \ref{fig:mnistlatent}, which shows the comparison between (a) the classic approach and (b) the quantum-based approach. 
The differences here can be appreciated by noticing cluster separation, within class distributions, and cluster overlaps. The figure suggests that the learned spaces are rich and discriminative as they appear, and if they are further re-trained separately for classification, it might yield very good results, even if the latent space is only two-dimensions. However, for a 64-dimensional space there are clear improvements in regards to cluster features, as seen in Fig. \ref{fig:mnistlatent} (c) and (d) for classic and quantum-based, respectively. Note that (c) and (d) are created with UMAP \cite{mcinnes2018umap}, which finds a two-dimensional representation of the $\mathbb{R}^{64}$ space.

\begin{figure}[ht!]\centering
    \begin{tabular}{cc}
\includegraphics[width=0.5\textwidth]
{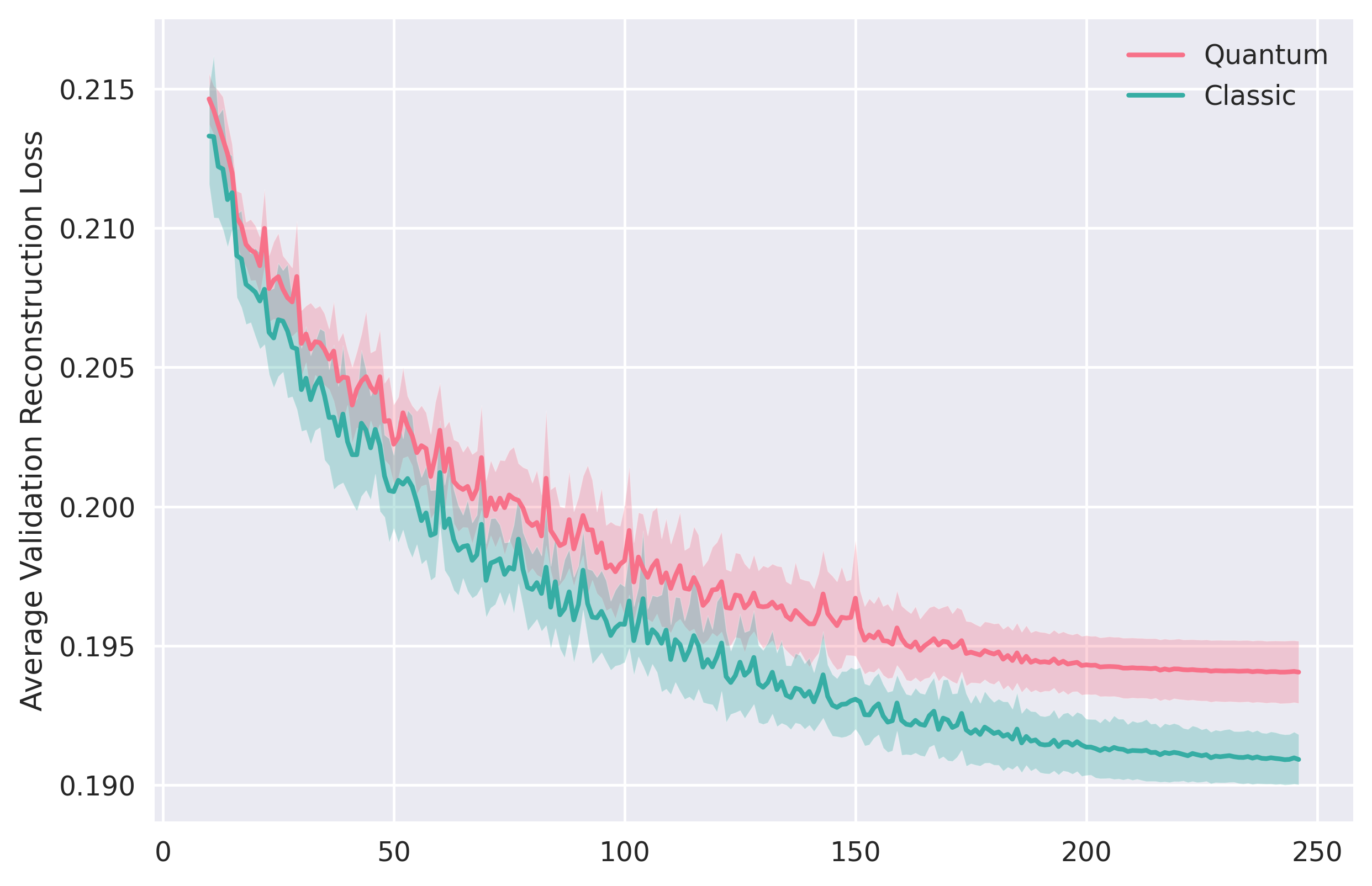} &
\includegraphics[width=0.5\textwidth]{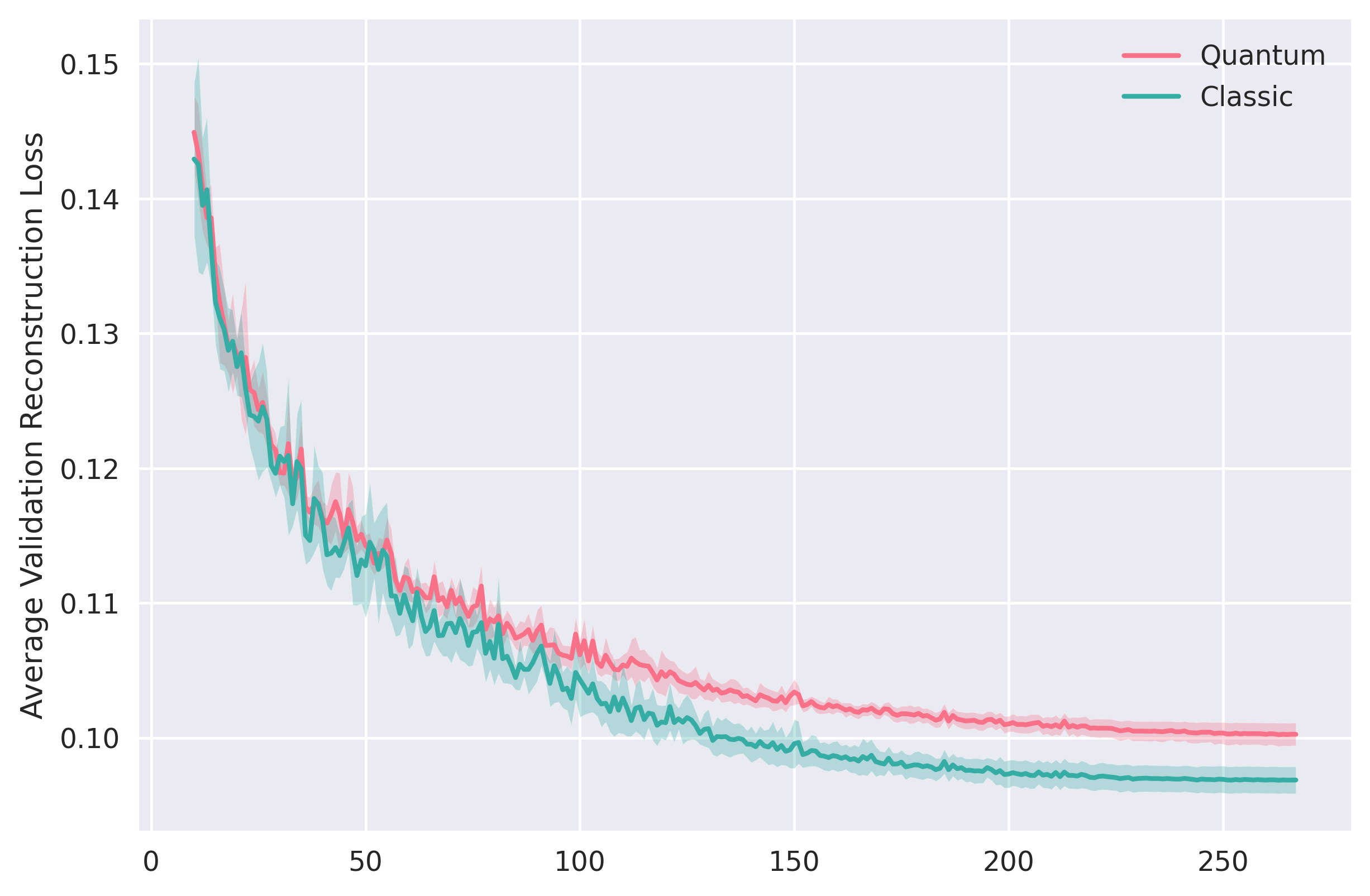}\\ 
{\tiny Epochs} &
{\tiny Epochs} \\ 
(a) &(b) \\
    \end{tabular}
    \caption{Average loss during gradient descent over a reconstruction loss on MNIST data. (a) is for $\mathbf{z} \in \mathbb{R}^{2}$ (b) is for $\mathbf{z} \in \mathbb{R}^{64}$.}
    \label{fig:mnistlosses}
\end{figure}

Finally, another objective way to compare performances is to observe the average behavior of the learning across epochs. Fig. \ref{fig:mnistlosses} shows the results of the two experiments with MNIST. 
In (a) we can compare the results using the two-dimensional latent space reconstructions. It shows that both methodologies are comparable and converge to a minimum without being statistically different, i.e., the standard deviation of both overlap most of the time. These results are produced with ten ($10$) randomized runs of the same experiment choosing random initial weights every time.
However, in (b) we can observe that the gap in performance between the two methodologies is statistically significant. What we can observe here is that both methodologies converge to different minima with nearly a $5\times10^{-3}$ difference in magnitude of the loss. The loss function used in all of our experiments is the \emph{binary cross entropy} loss with the \emph{adam} optimizer \cite{kingma2014adam}.

\subsection{CIFAR-10}

\begin{figure}[t!]\centering
    \begin{tabular}{c}
        \includegraphics[width=0.45\textwidth]{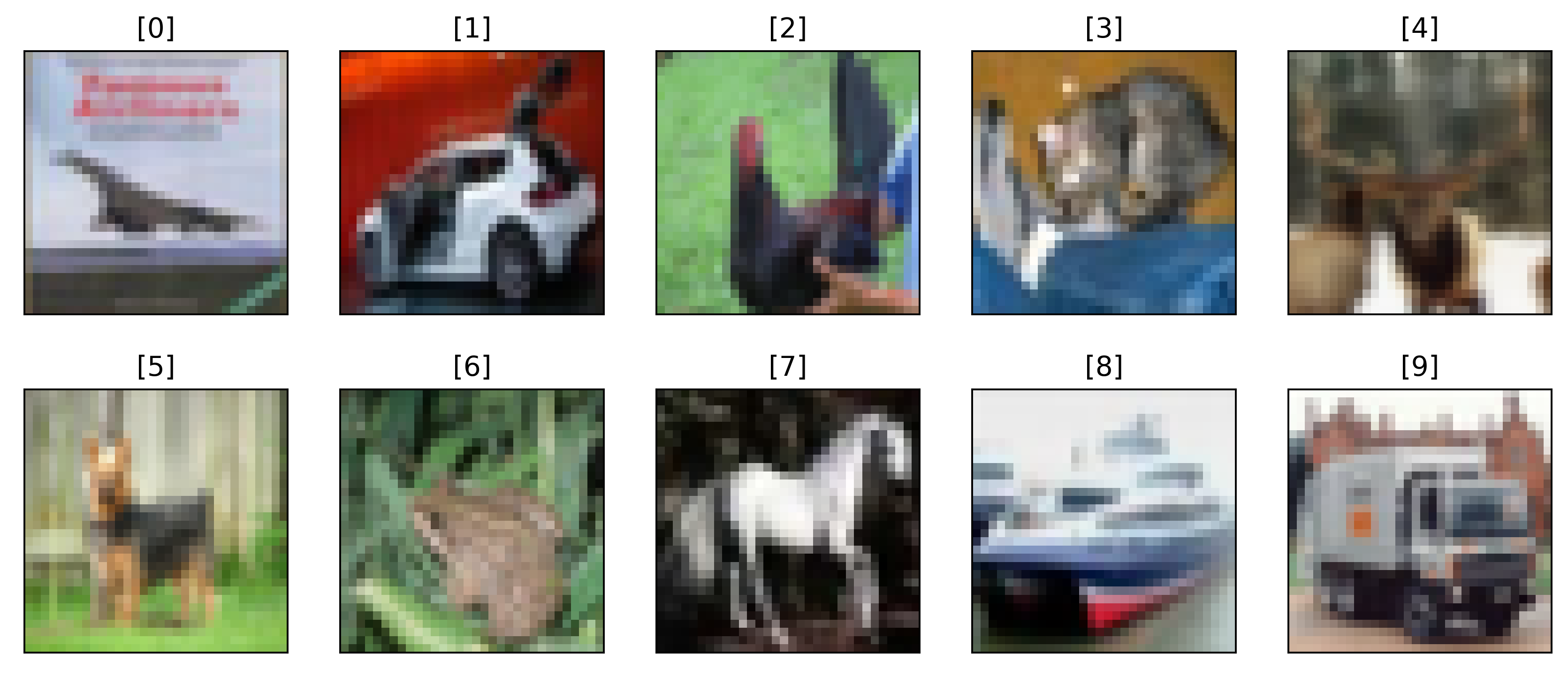} \\ (a) 
    \end{tabular}
    \begin{tabular}{cc}
        \includegraphics[width=0.45\textwidth]{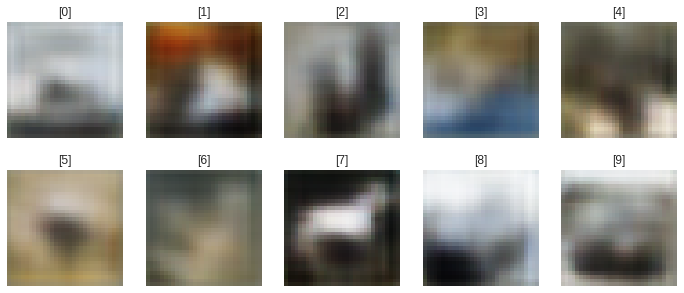} &
        \includegraphics[width=0.45\textwidth]{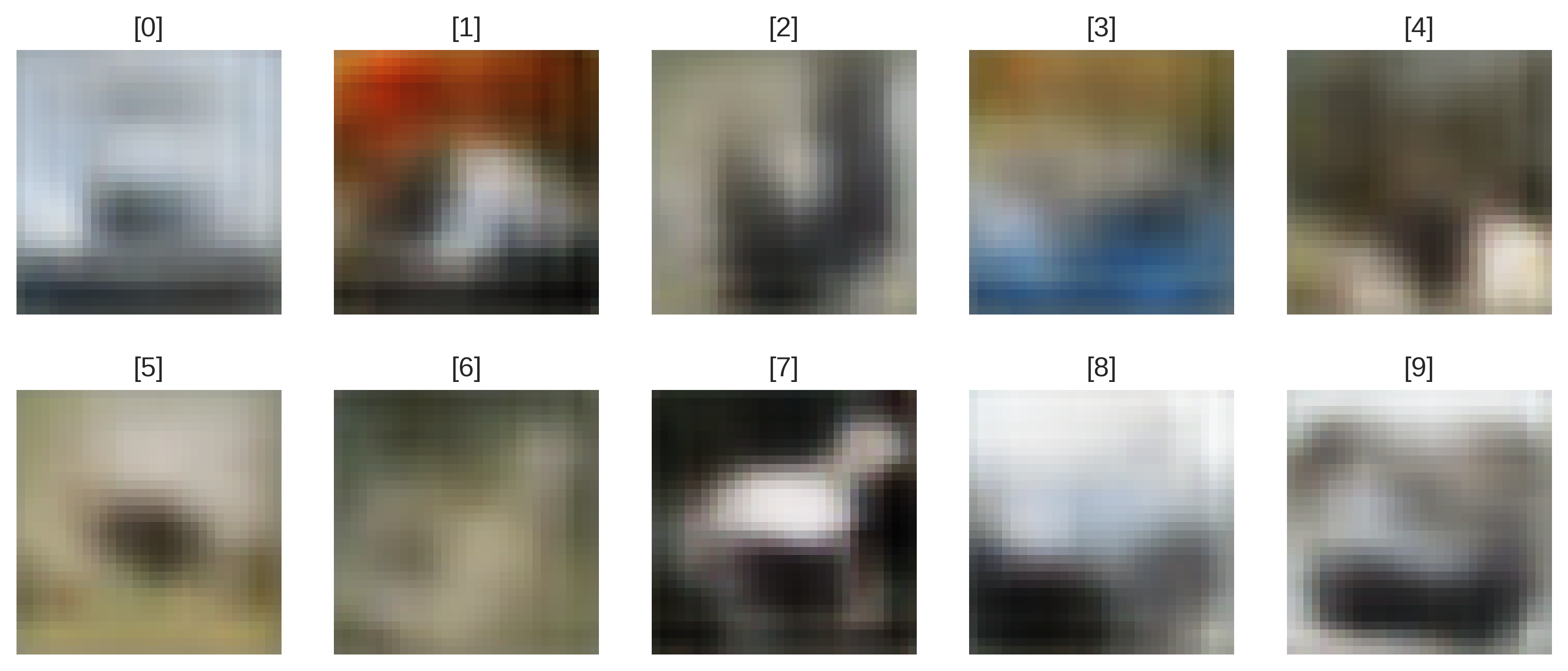} \\ (b) & (c)  
    \end{tabular}
    \caption{CIFAR-10 dataset samples and reconstruction. (a) Original samples. (b) Images reconstructed with the classic approach. (c) Images reconstructed with the proposed quantum approach.}
    \label{fig:cifar10}
\end{figure}

The dataset known as CIFAR-10 \cite{krizhevsky2009learning} consists of 50,000 color images of size $32 \times 32 \times 3$ pixels. These images correspond to 10 different categories and are labeled; however, the labels are not used during training. A sample of these pictures is shown in Fig. \ref{fig:cifar10} (a). The convolutional network for this dataset has the architecture shown in Table \ref{tab:cifararq}.
\begin{table}[ht!]\centering
    \caption{Configuration of the classic convolutional neural network for CIFAR-10. The layers marked with $\dag$ are the ones replaced by the quantum circuit convolutions. Note that for the dense layer marked with $\ast$ we also experimented with two neurons instead of 128 while the rest of the architecture remains the same.}
    \label{tab:cifararq}
    \begin{tabular}{|l|l|l|l|l|l|} 
    \hline
    Operation or Layer & Filters & Filter Size & Stride & Padding & Size of Output \\ \hline 
    $^\dag$Input image & - & - & - & - & \(32 \times 32 \times 3\) \\ \hline 
    $^\dag$Convolution &  48 & \(4 \times 4\) & \(1 \times 1\) & same & \(32 \times 32 \times 46\) \\  
    $^\dag$ReLU & - & - & - & - & \(32 \times 32 \times 48\) \\ \hline 
    $^\dag$Max pooling & - & \(4 \times 4\) & \(4 \times 4\) & same & \(8 \times 8 \times 48\) \\ \hline
    Convolution &  24 & \(3 \times 3\) & \(1 \times 1\) & same & \(8 \times 8 \times 24\) \\  
    ReLU & - & - & - & - & \(8 \times 8 \times 24\) \\ \hline 
    Convolution &  12 & \(2 \times 2\) & \(1 \times 1\) & same & \(8 \times 8 \times 12\) \\  
    ReLU & - & - & - & - & \(8 \times 8 \times 12\) \\ \hline 
    Max pooling & - & \(2 \times 2\) & \(2 \times 2\) & same & \(4 \times 4 \times 12\) \\ \hline
    Flatenning & - & - & - & - & 192 \\ 
    Dropout 20\% & - & - & - & - & 192 \\ 
    Dense (linear) & - & - & - & - & 192 \\ \hline
    $^\ast$Dense (tanh) & - & - & - & - & 128 \\ \hline
    Dense (linear) & - & - & - & - & 192 \\ 
    Reshape & - & - & - & - & \(4 \times 4 \times 12\) \\ \hline
    Convolution &  12 & \(2 \times 2\) & \(1 \times 1\) & same & \(4 \times 4 \times 12\) \\  
    ReLU & - & - & - & - & \(4 \times 4 \times 12\) \\ \hline 
    Up Sampling & - & \(2 \times 2\) & \(2 \times 2\) & same & \(8 \times 8 \times 12\) \\ \hline
    Convolution &  24 & \(3 \times 3\) & \(1 \times 1\) & same & \(8 \times 8 \times 24\) \\  
    ReLU & - & - & - & - & \(8 \times 8 \times 24\) \\ \hline 
    Convolution &  48 & \(1 \times 1\) & \(1 \times 1\) & same & \(8 \times 8 \times 48\) \\  
    ReLU & - & - & - & - & \(8 \times 8 \times 48\) \\ \hline 
    Up Sampling & - & \(4 \times 4\) & \(2 \times 2\) & same & \(32 \times 32 \times 48\) \\ \hline
    Convolution &  3 & \(4 \times 4\) & \(1 \times 1\) & none & \(32 \times 32 \times 3\) \\  
    Sigmoid & - & - & - & - & \(32 \times 32 \times 3\) \\ \hline 
    \end{tabular}
\end{table}
From the table, we can see that the first four elements of the table are marked with $^\dag$, indicating that those layers are the ones replaced with the quantum-based based methodology to replace traditional convolutions. These quantum convolutions start with the same single input image of $28 \times 28 \times 1$ and end up in multiple convolved images of size $7 \times 7 \times 16$. This makes the comparison between the classic convolutional AE approach and the quantum-based one, with the exception, of course, of the fact that in the classic approach, all convolutional layers are trainable. In contrast, the quantum-circuit convolutions, shown in Fig. \ref{fig:15Q}, are fixed. These facts related to the architecture are depicted in Fig. \ref{fig:architecture}.

\begin{figure}[ht!]\centering
    \begin{tabular}{cc}
        \includegraphics[width=0.5\textwidth]{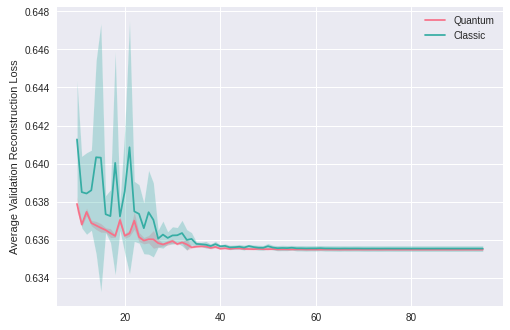} &
\includegraphics[width=0.5\textwidth]{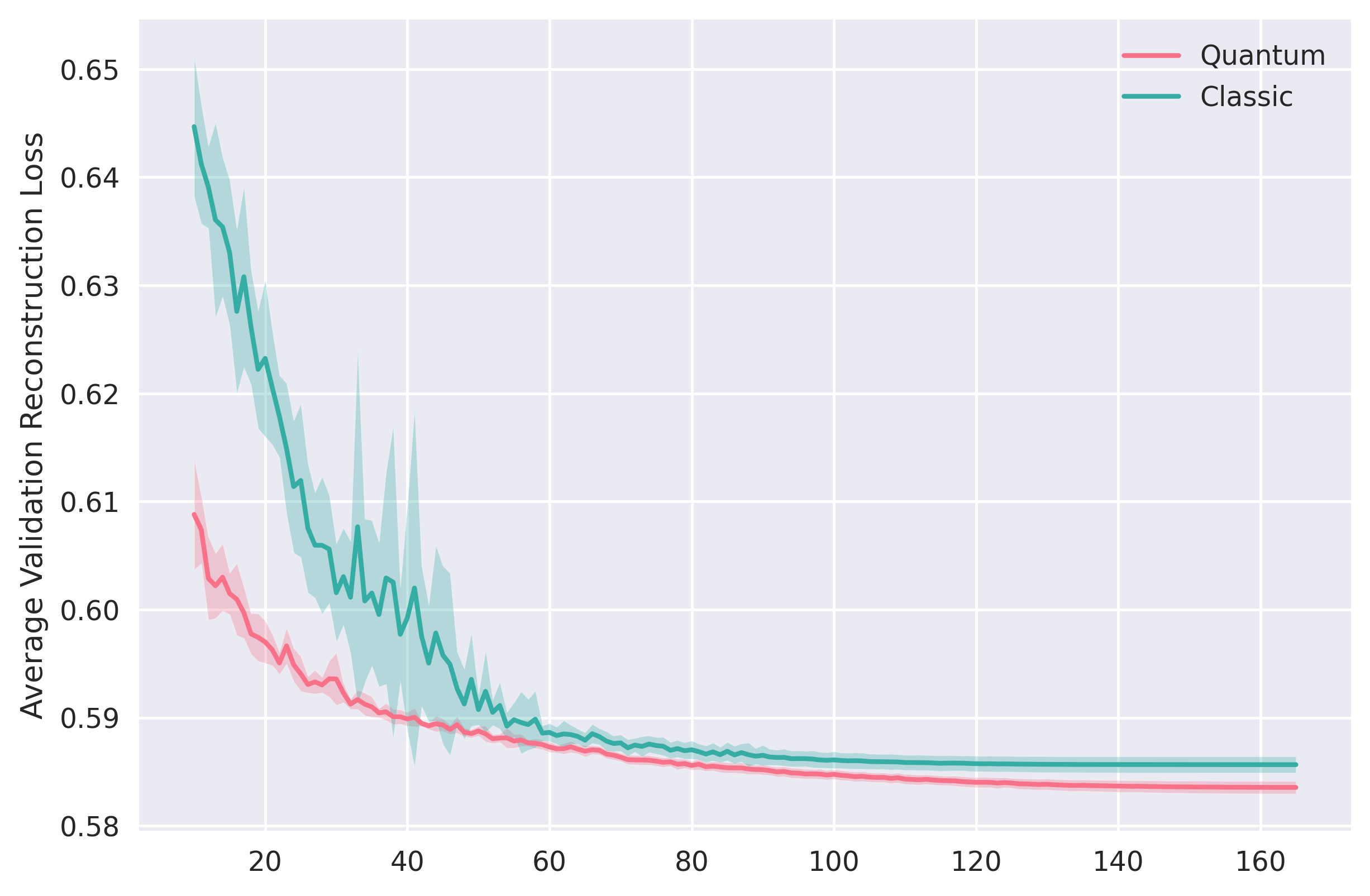}\\ {\tiny Epochs} & {\tiny Epochs}\\ 
(a) & (b) \\
    \end{tabular}
    \caption{Loss function minimization across different number of epochs. (a) is for a two-dimensional embedding space and (b) is for a 128-dimensional space.}
    \label{fig:cifarlosses}
\end{figure}

Similar to the experiments performed over MNIST in the previous section, we executed the same tests of performance, obtaining the reconstruction results shown in Fig. \ref{fig:cifar10}. In (b) we can see the reconstruction results of the classic autoencoder, and in (c) we see the corresponding quantum-based implementation. The results are comparable in terms of a visual inspection. However, when we display the average behavior of the optimization process, we can encounter interesting results, see Fig. \ref{fig:cifarlosses}.
From the figure, we observe that for an embedding space in $\mathbb{R}^2$, Fig. \ref{fig:cifarlosses} (a), the loss function minimization is comparable for both models and appears unstable at earlier iterations. However, for the larger embedding space in $\mathbb{R}^{128}$, Fig. \ref{fig:cifarlosses} (b), the quantum-based approach exhibits strong stability in earlier iterations in contrast to the classic approach. Both models eventually converge to a similar solution after a number of iterations.

\section{Discussion and Conclusions\label{sec:dis}}
The experiments performed here show that the proposed work is comparable to the classic approach in terms of reconstruction ability and convergence. However, in the case of a general-purpose image dataset such as CIFAR-10, the quantum-based approach offered early learning stability in learning higher-dimensional latent spaces, in contrast to the classic approach that converges slowly at first.

This paper presented an AE application with a particular quantum implementation in the algorithm in the form of quantum convolutions. We call this implementation a Quanvolutional autoencoder.
Our deployment shows the potential applications of quantum algorithms in learning image representations.

\section*{Acknowledgements}
The authors thank the Department of Computer Science at Baylor University for their support. This research was also funded, in part by the Baylor AI lab.

\bibliography{example_paper}

\end{document}